\documentclass[12pt]{article}
\usepackage[english]{babel}
\usepackage{graphicx,epsfig}
\makeindex \textwidth 170mm \textheight 220mm \topmargin -5mm
\newcommand{\be}{\begin{equation}}
\newcommand{\ee}{\end{equation}}
\newcommand{\bea}{\begin{eqnarray}}
\newcommand{\eea}{\end{eqnarray}}

\def\rfr#1{eq.(\ref{#1})}

\def\bb{\bibitem}
\def\eqi{\begin{equation}}
\def\eqf{\end{equation}}
\def\eqia{\begin{eqnarray}}
\def\eqfa{\end{eqnarray}}
\def\rp#1#2{{#1\over#2}}
\def\gr{General Relativity}

\def\lb#1{\label{#1}}
\def\grc{gravitomagnetic clock effect}

\begin{document}
\begin{titlepage}
\begin{flushright}
\today\\
BARI-TH/00\\
\end{flushright}
\vspace{.5cm}
\begin{center}
{\LARGE Testing the gravitomagnetic clock effect on the Earth with
neutron interferometry } \vspace{1.0cm}
\quad\\
{Lorenzo Iorio$^{*}$\\
\vspace{1.0cm}
Lorenzo.Iorio@ba.infn.it\\
\vspace{1.0cm}
\quad\\
{*}Dipartimento di Fisica dell' Universit{\`{a}} di Bari, via
Amendola 173, 70126, Bari, Italy\\} \vspace{1.0cm}

{\bf Abstract\\}
\end{center}

{\noindent \small The general relativistic gravitomagnetic clock
effect consists in the fact that two point particles orbiting a
central spinning object along identical, circular equatorial
geodesic paths, but in opposite directions, exhibit a time
difference in describing a full revolution. It turns out that the
particle rotating in the same sense of the central body is slower
than the particle rotating in the opposite sense. In this paper it
is proposed to measure such effect in an Earth laboratory
experiment involving interferometry of slow neutrons. With a
sphere of 2.5 cm radius and spinning at $4.3\times 10^4$ rad/s as
central source, and using neutrons with wavelength of 1 \AA\ it
should be possible to obtain, for a given sense of rotation of the
central source, a phase shift of 0.18 rad, well within the
experimental sensitivity. By reversing the sense of rotation of
the central body it should be possible to obtain a 0.06 fringe
shift.} \end{titlepage}
\newpage \pagestyle{myheadings} \setcounter{page}{1}
\vspace{0.2cm} \baselineskip 14pt

\setcounter{footnote}{0}
\setlength{\baselineskip}{1.5\baselineskip}
\renewcommand{\theequation}{\mbox{$\arabic{equation}$}}
\noindent

\section{Introduction}
One of the most intriguing consequences of Einstein's \gr\ is the
structure of the spacetime around a massive rotating body. The set
of its features is named gravitomagnetism [{\it Ciufolini and
Wheeler,} 1995; {\it Mashhoon et al.,} 2001] since  the
relativistic equations of motion of a test particle orbiting the
central spinning body, in the weak-field and slow-motion
approximation, are formally analogous to those of an electrically
charged particle acted upon electric and magnetic fields via the
Lorentz force.

Some of the consequences of the gravitomagnetic field are:\\
\textsl{a)} The precession of a gyroscope which will be tested in
the field of the Earth by the space mission GP-B [{\it Everitt et
al.,} 2001], scheduled to fly
in 2002, at a claimed precision level of 1$\%$\\
\textsl{b)} The dragging of the local inertial frames which is
currently measured in the field of the Earth as well by analyzing
the orbits of the laser-ranged passive geodetic satellites LAGEOS
and LAGEOS II [{\it Ciufolini et al.,} 2000]. To date, the
accuracy
amounts to $20\%$ \\
\textsl{c)} The \grc\ [{\it Mashhoon et al.,} 1999] consisting in
the fact that two counter-orbiting test particles placed on two
identical, circular equatorial geodesic orbits around a central
spinning mass take different times to complete a full revolution.
It turns out that the particle rotating in the same sense of the
central body is slower than the particle moving in the opposite
sense. At present, the feasibility of a space-based experiment
aimed at the detection of this effect in the terrestrial field is
under consideration by the scientific community [{\it Gronwald et
al.,} 1997;  {\it Lichtenegger et al.,} 2000; 2001; {\it Iorio,}
2001a; 2001b]

Such relativistic effects are very tiny and their detection in the
terrestrial environment via space-based missions is a very
demanding task due to many other competing forces acting upon the
satellites to be employed which may alias the recovery of the
relativistic feature of interest. Last but not least, these kinds
of experiments are, obviously, very expensive.

In this paper we will focus on the \grc\ and on the possibility of
measuring it on the Earth in a laboratory experiment involving
neutron interferometry [\textit{Rauch and Werner,} 2000]. Such
technique has proved itself useful in the so called COW experiments which
detected an
interference path due to the gravitoelectric field of the Earth
[{\it Overhauser and Colella,} 1974; {\it Colella et al.,} 1975;
\textit{Staudenmann et al.,} 1980; \textit{van der Zouw et al.,}
2000]. In [{\it
Cohen and Mashhoon,} 1993] it can be found a preliminary discussion of
an attempt of detecting the gravitomagnetic field of the Earth in a
COW-like interference experiment. A time-shift due to gravitomagnetism in
the field
of the quantum domain can be found in [{\it Ahluwalia}, 1997]; see
also  [{\it Mashhoon et al.,} 2000; \textit{Mashhoon}, 2000]
for the role of gravitomagnetism in quantum theory. For a
review of proposed terrestrial experiments aimed at the detection of some
other gravitomagnetic features see chapter 6.9 in [{\it Ciufolini
and Wheeler,} 1995].

The paper is organized as follows: in Section 2 we first review a recent
proposal for detecting the gravitomagnetic time shift of a couple of
electromagnetic orthogonally running waves; subsequently, we describe our
approach with neutrons. Section 3 is devoted to a brief discussion on the
validity of the approach followed while in Section 4 we summarize the
conclusions.
\section{Proposal of an actual laboratory experiment }
\subsection{Electromagnetic interferometry}
In [{\it Tartaglia and Ruggiero,} 2001] the authors work out, in
the weak-field and slow-motion approximation, a gravitomagnetic
time difference for two electromagnetic waves one running along an
equatorial circular path and the other one along a polar one
around a central massive spinning body. Subsequently, they propose
to detect it in a laboratory experiment on the Earth by using as
central source a hollow spherical rotating shell and by means of
two fixed orthogonal wave guides in order to make the waves to
interfere. The investigated time shift is given by \eqi \Delta
T=\frac{\pi}{2}(\frac{J}{M})^2\frac{1}{c^{3}R_l}=\frac{2}{9}\frac{\pi}{c^{3}}\frac{R^{2}}{R_l}\frac{\sigma_{m}}{\rho}
,\lb{tart}\eqf in which $J$ is the spin of the central rotating
body, $M$ is its mass, $c$ is the speed of light \textit{in
vacuum}, $R$ is the radius of the shell, $R_l$ is the radius of
the light's path, $\rho$ is the material density of the shell and
$\sigma_{m}$ is the allowable resistance of the shell's material.
Note that such an effect is very tiny because it is of order
$\mathcal{O}(c^{-3})$. Indeed, for $\sigma_{m}=2,000$ MPa,
$\rho=1,700$ kg m$^{-3}$ and $R=1$ m the time shift amounts to
$\Delta T=3\times 10^{-20}$ s only. By using visible light with
wavelength of, say, $\lambda=5\times 10^{-7}$ m it yields a
relative phase shift\eqi\Delta\Phi=2\pi \frac{c}{\lambda}\Delta
T\sim 10^{-5}\ {\rm rad},\eqf with a beam intensity relative
change at the interference of only\eqi\delta
I/I=\frac{1}{2}(1-\cos\Delta\Phi)\sim 10^{-9}.\eqf
\subsection{Neutron interferometry}
Instead of considering the gravitomagnetic time difference for two
electromagnetic waves running along orthogonal paths, for which
$\Delta T\sim\mathcal{O}(c^{-3})$ only, we propose to adopt two
counter-orbiting, circular beams of slow neutrons running in the
equatorial plane of the central body so to adopt the well known
[{\it Mashhoon et al.,} 1999] time shift\eqi\Delta
T=T_{+}-T_{-}=4\pi\frac{J}{Mc^{2}},\lb{mash}\eqf in which $T_{+}$
is the period of the particle rotating in the same sense of the
central body and $T_{-}$ is the period of the particle rotating in
the opposite sense of the central body: as it can be seen, the
co-rotating particle is slower than the counter-rotating one. It
is worth noting that \rfr{mash}, which has been derived in the
weak-field and slow-motion approximation of \gr\ as well, is of
order $\mathcal{O}(c^{-2})$. Moreover, \rfr{mash}, contrary to
\rfr{tart}, is independent of the radius of the particles' orbits
and depends entirely on the characteristics of the central body.
 Note also that
the time shift amount is independent of the gravitational constant
$G$ and of the mass of the central object if it is endowed with
suitable symmetries; conversely, it depends linearly on its
angular speed $\Omega$ and quadratically on its radius $R$:
indeed, for a sphere \rfr{mash} becomes\eqi\Delta T=
4\pi\frac{I\Omega}{c^{2}}=\frac{8\pi}{5}\frac{R^{2}\Omega}{c^{2}},\eqf
where $I$ is the momentum of inertia of the sphere.  The maximum
angular speed is fixed by the material properties of the object:
\eqi \Omega_{\rm max}=\sqrt{\frac{\sigma_{\rm max}}{\rho
R^{2}}},\eqf where $\sigma_{\rm max}$ is the maximum attainable
stress at the equator (unidimensional stresses are assumed) and
$\rho$ is its mass density [{\it Tartaglia and Ruggiero,} 2001].
The radius $R$ is limited by the size of the experimental setup:
for example, in interferometry experiments with thermal neutrons
the typical dimensions of the interferometer are of the order of
$10^{-1}$ m, so that $R=2.5\times 10^{-2}$ m seems a reasonable
value.  With such value for the radius and with $\sigma_{\rm
max}=2\times 10^{9}$ Pa and $\rho=1700$ Kg m$^{-3}$ [{\it
Tartaglia and Ruggiero,} 2001] the maximum attainable angular
speed is $\Omega_{\rm max}=4.33\times 10^{4}$ rad s$^{-1}$.  So,
by using a
 small sphere of $2.5\times 10^{-2}$ m radius spinning at
$\Omega=4.33\times 10^{4}$ rad s$^{-1}$ \rfr{mash} yields $\Delta
T=1.51\times 10^{-15}$ s which is five orders of magnitude larger
than the electromagnetic time shift worked out in [{\it Tartaglia
and Ruggiero,} 2001]. If we use neutrons with De Broglie's
wavelength $\lambda_{\rm n}=h/p_{\rm n}=h/m_{\rm n} v_{\rm n}
=1\times 10^{-10}$ m $\equiv$ 1\AA, where $h$ is the Planck's
constant, we can be successful in raising  the relative phase
shift to the sensitivity limit of $10^{-2}-10^{-3}$ rad. Indeed,
after a full revolution around the spinning sphere, for a given
sense of its rotation, the two neutron beams would accumulate a
phase shift at the interference point\eqi\Delta\Phi^{\rm
neutron}_{\rm grav}=2\pi\frac{v_{\rm n}}{\lambda_{\rm n}}\Delta
T_{\pi}=2\pi\frac{m_{\rm n} v_{\rm n}^2}{h}\Delta
T_{\pi}=4\pi^{2}\frac{m_{\rm n}}{hM}(\frac{v_{\rm
n}}{c})^{2}J.\lb{dfg}\eqf By considering thermal neutrons with
wavelength of 1\AA\ we would have $v_{\rm n}=3.9\times 10^3$ m
s$^{-1}$ and the phase shift would amount to $\Delta\Phi^{\rm
neutron}_{\rm grav}=1.88\times 10^{-1}$ rad.

Moreover, by repeating the experiment inverting the sense of
rotation of the central sphere the neutron beams would experience
a time shift $\Delta T^{'}=-\Delta T$ since the beam which was
formerly slower now would become faster and vice versa. So it
would be possible to observe a fringe shift of\eqi\Delta
N=\frac{v_{\rm n}}{\lambda_{\rm n}}(\Delta T_{\pi}-\Delta
T^{'}_{\pi})=2\frac{v_{\rm n}}{\lambda_{\rm n}}\Delta
T_{\pi}=6\times 10^{-2}\eqf for $\lambda_{\rm n}=1\times 10^{-10}$
m. The minimum appreciable fringe shift amounts to $10^{-3}$. It
is interesting to note that \rfr{dfg} is not independent of the
mass of neutron. Moreover, it should be pointed out that in our
setup, contrary to the COW experiments, the single crystal
interferometer would remain always horizontal, neither it would be
rotated along a given axis so that the gravitoelectric field of
the Earth would not influence the outcome of the experiment.

Note that by using ultracold neutrons with, e.g., $\lambda_{\rm
n}=7\times 10^{-8}$ m and $v_{\rm n}=5.7$ m s$^{-1}$, we would
obtain, from one hand, larger geometries of the experimental setup
would be allowable,  but from the other hand the expected phase
shift would fall well below the experimental sensitivity which, in
this case, is smaller than that of thermal neutrons
interferometry.
\section{Discussion}
Here we will show that we can apply \rfr{dfg} and \rfr{mash} to
our scenario.

Regarding the expression used for the phase shift, let us recall
that neutron interferometry, in general, is most appropriately
described by taking the Hamiltonian
$H=\hat{\textbf{p}}^{2}/2m+mU(\textbf{r})$  and performing a WKB
approximation giving a Hamilton-Jacobi equation which can be
solved for the modulus of the momentum
$p(\textbf{r})=p\sqrt{1+2m^{2}U(\textbf{r})/p^{2}}$ where $p$ is
the momentum of the neutrons at the beam splitter
[\textit{L\"{a}mmerzahl,} 1996] and $U$ is the gravitational
potential acting upon the particles (In fact, it should be
considered the nuclear potential as well [{\it Mashhoon,} 1988],
but it acts only over very short distances and only at the
reflections at the crystal lattice of the mirrors). In our case
$U(\textbf{r})_{\rm GM}=-\frac{(\textbf{A}_g\cdot\textbf{v})}{c}$
in which $\textbf{A}_g$ is the gravitomagnetic vector potential
given by \eqi {\bf A}_g=-\rp{2G}{c}\rp{{\bf J}\times{\bf
r}}{r^3}.\eqf For paths in the equatorial plane of the central
source \eqi U_{\rm GM}=2\frac{GJ}{c^{2}}\frac{v}{r^2}.\eqf For the
geometry of our setup ($J=1.2$ Kg m$^2$ s$^{-1}$, $v_{\rm
n}=3.95\times 10^{3}$ m s$^{-1}$, $r_{\rm n}=5\times 10^{-2}$ m )
we have: $2m_{\rm n}^2U_{\rm GM}/p^2=3.59\times 10^{-28}$, so that
$p(\textbf{r})=p$. In this limit the phase shift, which in general
is given by \eqi \Delta\Phi^{\rm neutron}_{\rm
grav}=\frac{1}{\hbar}\oint\textbf{p}(\textbf{r})\cdot\textbf{dr},\eqf
becomes just \rfr{dfg}.

Let us recall that \rfr{mash} has been worked out in the general
relativistic weak-field and slow-motion approximation for a couple
of massive particles without internal degrees of freedom following
well defined closed geodesic circular paths. Regarding the latter
point, because the size of the neutron wave packet can be assumed
much more smaller than the macroscopic dimension of the loop
formed by the two alternate paths, we can apply the concept of a
classical trajectory. Of course, the slow neutrons fly at
nonrelativistic speeds and there is no question about the weakness
of the gravitomagnetic field generated by the spinning sphere. A
major objection could be that \rfr{mash} has been derived for
geodesic orbits [{\it Tartaglia and Ruggiero,} 2001b], while in
our proposed experiment there are  the short range nuclear forces
at the momentum-conserving reflections [\textit{L\"{a}mmerzahl},
1996] on the crystal mirrors to be accounted for as well. They
would introduce non-geodesic contributions to the neutrons'times
which could invalidate the use of \rfr{mash}. Regarding this
point, it could be noted that, for a given sense of flight, the
opposite one is the time-reversal $\mathcal{T}$ image of the
former one. The nuclear interactions between the neutrons and the
crystal lattice's atoms are invariant with respect $\mathcal{T}$.
Then, it could be argued that the non-geodesic contributions to
the neutrons'periods $T=T_{\rm geod}+T_{non geod}$ are equal in
value and sign so that they cancel out in the difference leaving
only the geodesic part given by \rfr{mash}.

It may be useful also to stress that \rfr{mash} is the time
difference as seen by distant inertial observers (or by local
non-rotating observers) [{\it Tartaglia,} 2000]. In our setup the
laboratory frame is rotating with the Earth (see [{\it Tartaglia
and Ruggiero,} 2001b] for interesting critical remarks), but, due
to the extension of the setup in space and time (the lifetime of
neutrons is almost 11 min.) and its geometry, for all practical
purpose the adopted frame can be considered as inertial: recall
that the proposed angular speed of the central sphere is
$\Omega\simeq 10^4$ rad s$^{-1}$, while the Earth's proper angular
speed is $\Omega_{\oplus}\simeq 10^{-5}$ {\rm rad} s$^{-1}$. This
suggests that the Sagnac effect should be negligible in this case.

Concerning the fact that neutrons are $1/2$ spin particles, under
the action of the non-uniform gravitomagnetic field of the
rotating body their beams could experience gravitomagnetic
Stern-Gerlach forces [{\it Mashhoon,} 2000]\eqi\textbf{F}_{\rm
SG}=\frac{3GJ}{c^{2}r^{4}}\{[5(\overrightarrow{\sigma}\cdot\hat{\textbf{r}})
(\hat{\textbf{J}}\cdot\hat{\textbf{r}})-\overrightarrow{\sigma}\cdot
\hat{\textbf{J}}]-(\overrightarrow{\sigma}\cdot\hat{\textbf{r}})\hat{\textbf{J}}-(\hat{\textbf{J}}
\cdot\hat{\textbf{r}})\overrightarrow{\sigma}\}, \lb{stg}\eqf
where  $\overrightarrow{\sigma}$ is the particle's spin vector
($\sigma=\hbar s$) and $r$ is the particle's position vector. Note
that \rfr{stg} is independent of the neutron mass. For an orbit
radius of $r\geq 5\times 10^{-2}$ m and for the sphere  considered
here ($J=1.2$ Kg m$^2$ s$^{-1}$) the amplitude of \rfr{stg}
amounts to $4.4\times 10^{-56}$ N only, so that we can completely
neglect it and deal with neutrons as test particles.

\section{Conclusions}
Here we have explored the possibility of setting an Earth
laboratory experiment with neutron interferometry aimed at the
detection of the phase shift which would be induced on the paths
of two coherent beams of slow neutrons by the general relativistic
gravitomagnetic field of a small, rapidly spinning sphere. For a
given sense of rotation of the latter, the neutrons' phase shift
amounts to 0.18 rad. By reversing the sense of rotation of the
central source it should be possible to obtain a fringe shift in
the interference pattern of 0.06. These results hold for a sphere
of 2.5 cm radius spinning at $4.3\times 10^4$ rad s$^{-1}$ and for
neutrons with a wavelength of 1 \AA.

Of course, the practical implications of the realization of the proposed
experiment
must be analyzed in detail.
\section*{Acknowledgements} I warmly thank S. Pascazio and P. Facchi for
their kind attention and fruitful discussions. I
would like to thank also L. Guerriero for
his encouragement and support and D. V. Ahluwalia for the interest.

\end{document}